\documentclass[aps,prb,superscriptaddress,twocolumn]{revtex4-1}
\usepackage{graphicx}% Include figure files
\usepackage{dcolumn}% Align table columns on decimal point
\usepackage{bm}% bold math
\usepackage{hyperref}
\usepackage{xcolor}

\definecolor{darkgreen}{rgb}{0.0,0.4,0.0}

\newcommand{\tc}{T$_{\textmd c }$}
\newcommand{\Tc}{T$_{\textmd c }$}

\newcommand{\af}{$\alpha^2F$}
\newcommand{\omlog}{$\omega_{\textrm {log}}$}

\newcommand{\MPIHalle}{Max-Planck-Institut f\"ur Microstruktur Physik, Weinberg 2, 06120 Halle, Germany} 
\newcommand{\Sapienza}{Dipartimento di Fisica, Università di Roma La Sapienza, Piazzale Aldo Moro 5, I-00185 Roma, Italy}
\begin{document}

\title{A possible explanation for the high superconducting \Tc\ in bcc Ti at high pressure}

\author{Antonio Sanna}
\email{sanna@mpi-halle.mpg.de}
\affiliation{\MPIHalle}

\author{Camilla Pellegrini} 
\email{pellegrini.physics@gmail.com}
\affiliation{\MPIHalle}

\author{Simone di Cataldo}

\affiliation{Institute of Solid State Physics, TU Wien, 1040 Vienna, Austria}
%\affiliation{Institute of Theoretical and Computational Physics, Graz University of Technology, NAWI Graz, 8010 Graz, Austria

\affiliation{\Sapienza}
\author{Gianni Profeta}
\affiliation{Dipartimento di Scienze Fisiche e Chimiche, Università degli Studi dell’Aquila, Via Vetoio 10, I-67100 L'Aquila, Italy} 
\affiliation{CNR-SPIN c/o Dipartimento di Scienze Fisiche e Chimiche, Università degli Studi dell’Aquila, Via Vetoio 10, I-67100 L’Aquila, Italy} 
\author{Lilia Boeri}
\email{lilia.boeri@uniroma1.it}
\affiliation{\Sapienza}
\affiliation{Centro Ricerche Enrico Fermi, Via Panisperna 89 A, 00184 Rome, Italy}

%%% Cite latest Yanming's paper, as a possible evidence of structural distortions

\date{\today}

\begin{abstract}
% Motivated by the recent observation of a 26~K superconducting transition in titanium at high pressure, an astonishing discovery for an elemental metal, we have performed accurate theoretical and computational study to find a rationale for this observation. We show that, according to first principles calculations, the expected critical temperature should be significantly lower than the experimentally observed. 
% To reconcile experimental and theoretical results, we have followed up the hypothesis of the possible presence of defects in the lattice, in particular of Ti vacancies. 
% These type of lattice defects cause a pressure dependent phonon softening and a significant increase of the electron -phonon coupling at high pressure, definitely reconciling theoretical predictions with the observed \Tc\ trend. 

Motivated by unexpected reports of a 26~K superconducting transition in elemental titanium at high pressure, we carry out an accurate \emph{ab initio} study of its properties to understand the rationale for this observation. The critical superconducting temperatures (\tc's) predicted under the assumption of a phononic pairing mechanism are found to be significantly lower than those experimentally observed. We argue that this disagreement cannot be explained by an unconventional coupling, as previously suggested, or by the existence of competing metastable structural phases. As a physically meaningful hypothesis to reconcile experimental and theoretical results, we assume the presence of Ti vacancies in the lattice. Our first-principles calculations indeed show that lattice vacancies can cause pressure dependent phonon softening and substantially increase the electron-phonon coupling at high pressure, yielding computed \tc's in agreement with the experimental measurements over the full pressure range from 150 to 300 GPa. We expect the proposed \tc\ enhancement mechanism to occur on a general basis in simple high-symmetry metals for various types of defects.

%We present an ab initio study of superconductivity in elemental titanium at high pressure.
%The calculated critical temperatures, as predicted assuming a phononic pairing mechanism, result to be significantly lower than those experimentally observed. 
%We argue that such disagreement can not be explained by an unconventional pairing coupling or by competing structural phases.
%To reconcile experimental and theoretical results we follow up the hypothesis of the possible presence of defects in the lattice, in particular of Ti vacancies. Our study shows that these type of lattice defects cause a pressure dependent phonon softening and a significant increase of the electron -phonon coupling at high pressure, leading to calculated critical temperature which are consistent with the experimental measurements in the entire pressure range from 150 to 300 GPa. 
\end{abstract}

\pacs{Valid PACS appear here}
\maketitle

\section{introduction}

\vspace{0.5cm}

%Most transition metals exhibit superconductivity at ambient pressure, with critical temperatures (\tc) ranging from a tenth of Kelvin to about 10 K. Superconductivity is typically enhanced by pressure, so that not only transition metals but a large majority of elemental materials become superconducting under pressure with \tc\ values which can be sensibly higher than 10~K~\cite{FloresBoeri_PerspectiveOnConvetionalHiTcSc_PhysRep2020,Buzea_PuzzleSCelements_SUST2005,Jinjun_Scanadium36K_PRB2023,Shimizu_Lipressure20K_Nature2002,Sakata_CaVII_29K_PRB2011,Shimizu_EHP2005}. 
%Scandium reaches \tc=36K at 260~GPa~\cite{Jinjun_Scanadium36K_PRB2023}, lithium has \tc$\sim$15K at 30GPa, \tc's exceeding 20 K reported also in Ca  .... 
While room pressure superconductivity measurements are \textit{usually} quite reliable, the interpretation of high-pressure phase diagrams of materials is often non trivial, due to intrinsic experimental limitations\cite{Shimizu_SCelementsPressure_PhysC2018}: samples are often small, pressure hysteresis is hard to control, {\em in-situ} measurements are obviously difficult, and the number of physical properties that can be independently measured is limited. In light of this, first-principles calculations based on density functional theory (DFT), play an essential role in high-pressure superconducting research. Indeed few people would nowadays object that DFT is the only framework which can consistently predict and explain structural, electronic, dynamical and superconducting properties of materials\cite{FloresBoeri_PerspectiveOnConvetionalHiTcSc_PhysRep2020}. As such, DFT-based calculations are becoming increasingly accepted as an independent source of validation for new high-pressure experimental discoveries. To give a few examples, reports of high-pressure superconductivity in sulfur\cite{DrozdovEremets_SH3_Nature2015}, phosphorus\cite{Drozdov_ph3_arxiv2015} and lanthanum hydrides\cite{Hemley-LaH10, Eremets-LaH10}, having been confirmed  by extensive and independent {\em ab initio} simulations\cite{Duan_SciRep2014, SH3-Antonio, PH3-ours, PH3-Zurek, LaH10-Errea}, are now considered established results.  On the other hand, the reports of room-temperature superconductivity in carbon sulphur hydride\cite{Dias-retracted} and, more recently, N-doped Lu-hydride,\cite{Dias_LUNH23,Ming_LUNH23}
which have not yet found convincing theoretical explanations\cite{Flores-Ryo_PRB2021,Boeri_LUNH23,Zurek_LUNH23,Errea_LUNH23,Goedecker_LUNH23}, are still considered controversial. %%% UPDATE REFS HERE!

This paper deals with another recent disputed case in superconducting research at high pressure:
the observation of a relatively high-\tc\ superconducting phase (\tc=26 K) in elemental titanium at high pressure (200-300 GPa), which has been reported independently by two groups\cite{Zhang2022_RecordHiTcElementSCinTi_NatComm2022, Liu_deltaPhase_Ti_200GPA_PhysRevB.105.224511}, yet explained in terms of conflicting pairing mechanisms based on different DFT results for the electron-phonon coupling. The thermodynamically stable phase of titanium in the 200-300 GPa pressure range is bcc ($Im-3m$). For this phase, Zhang and coworkers\cite{Zhang2022_RecordHiTcElementSCinTi_NatComm2022} have calculated a phonon-mediated \tc\ of $\simeq$ 5~K, thereby attributing the large discrepancy with the experimental data to unconventional pairing, experimental artifacts and/or other effects. 
On the other hand, Liu and collaborators~\cite{Liu_deltaPhase_Ti_200GPA_PhysRevB.105.224511} claim that their measurements can be explained within standard electron-phonon superconducting theory. However, the electron-phonon coupling constant they compute is so large that an extremely high value of the Coulomb pseudopotential $\mu^*$ must be introduced ad hoc in the McMillan formula so as to reproduce the experimental results.
Therefore, despite the consensus on the experimental measurements and the apparent simplicity of the theoretical framework, a clear understanding of the origin of the high temperature superconducting phase in titanium at high pressure is still missing.

In this work, using cutting-edge first principles DFT calculations, we independently investigate the superconducting state of high-pressure phases of titanium, in order to establish to what extent the experimental results can be understood within the frame of  conventional superconductivity. 
Our simulations prove that the theoretically predicted \tc's are indeed much lower than the experimental measurements, consistently with the calculations of Ref.~\onlinecite{Zhang2022_RecordHiTcElementSCinTi_NatComm2022}. We further show that an unconventional pairing mechanism, driven by an incipient magnetic instability, is incompatible with the calculated electronic properties and that there are no competing near-stability phases, which could justify the observed \Tc\ within conventional superconductivity. We propose and substantiate by first principles calculations a simple physical scenario in which the existence of structural defects (Ti vacancies) in the bcc lattice could account for the measured \Tc's by means of a conventional electron-phonon mechanism.

%Then, we will argue that an unconventional mechanism, driven by an incipient magnetic instability, is incompatible with the calculated electronic properties and that there are no competing, near-stability, phases which could justify the observed \Tc.

\section{Methods}
Normal-state properties have been computed within DFT~\cite{KohnSham_PR1965,HohenbergKohn_DFT_PR1964} as implemented in the Quantum Espresso code~\cite{QUANTUMESPRESSO} using the PBE-Sol approximation~\cite{PBEsol} for the exchange-correlation functional, which is generally more accurate than PBE to predict the lattice properties of crystalline materials. Atomic core states have been included in the norm-conserving pseudo-potential approximation,
using pseudoptentials from the pseudo Dojo distribution\cite{pseudoDojo_Site,VanSetten_PseudoDojo_CompPhysComm2018}. Valence states have been expanded in plane-waves with a kinetic energy cutoff of 100~Ry. Phonon frequencies and electron-phonon coupling matrix elements have been computed by means of density functional perturbation theory~\cite{Baroni_1987a,Baroni_DFPT_RMP2001}; different
$\mathbf{k}$- (electrons) and $\mathbf{q}$- (phonons) grids have been employed for reciprocal space integration, depending one the unit cell size (as listed in Sec.~\ref{sec:phases} and in the Supplemental Material~\cite{Supplemental_Material}), with a Gaussian smearing width of 0.02 Ryd.
Superconducting properties have been obtained within density functional theory for superconductors~\cite{OGK_SCDFT_PRL1988,Lueders_SCDFT_PRB2005,Marques_SCDFT_PRB2005}, using the functional form proposed in Ref.~\onlinecite{SPG_Functional_PRL2020}, where Coulomb interactions have been calculated in the random phase approximation as implemented in the Elk code~\cite{elk}. Further details are provided as Supplemental Information~\cite{Supplemental_Material}.

Crystal structure prediction calculations have been carried out using evolutionary algorithms as implemented in the Universal Structure Prediction Evolutionary X-stallograpy code (USPEX)\cite{GLASS2006713, LYAKHOV20131172}
employing a fixed unit cell of 8 atoms, an initial population size of 40 individuals for the first generation, and 20 individuals for the others, for a total of 20 generations. All generated crystal structures have been relaxed using a 5-step relaxation procedure with progressively tighter constraints, using the Vienna ab initio Simulation Package (VASP)\cite{VASP_Kresse}, with Projector Augmented Wave pseudopotentials in the Perdew-Burke-Ernzerhof approximation for the exchange-correlation functional.
The total enthalpy at the final step has been computed using $k$-space density of 0.25 $k$-points/\AA, a gaussian smearing  of 0.03 eV, and a cutoff energy of 600 eV. 

\section{Results}

\subsection{$\delta$ and $\beta$  phases of $\text{Ti}$}

\begin{figure}[t] 
\includegraphics[width=\columnwidth]{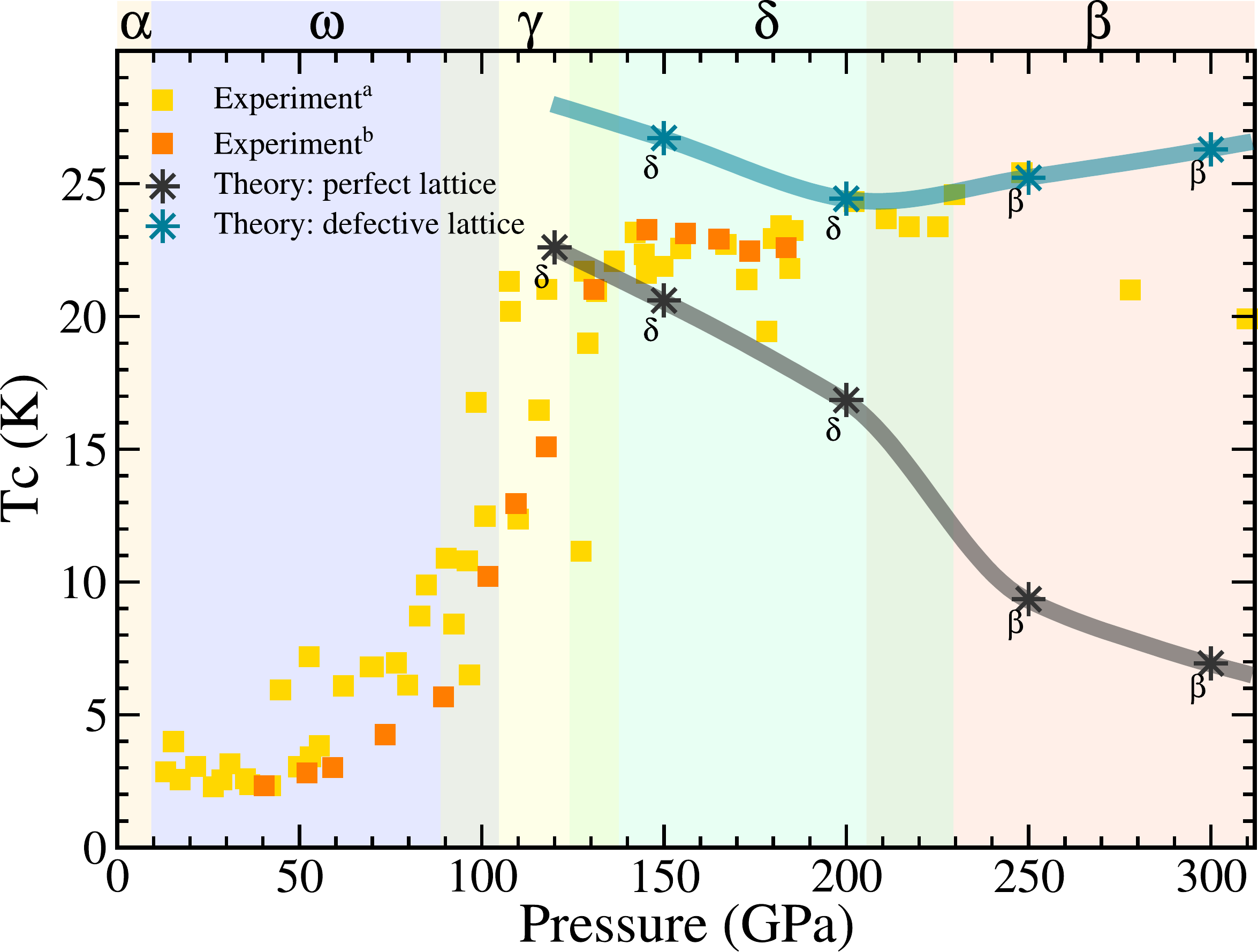} 
\caption{Comparison between theoretical and experimental superconducting critical temperatures of Ti at high pressure. Yellow and orange squares denote the experimental results from Refs.~\onlinecite{Zhang2022_RecordHiTcElementSCinTi_NatComm2022} and ~\onlinecite{Liu_deltaPhase_Ti_200GPA_PhysRevB.105.224511}. Black and blue stars represent fully \emph{ab initio} SCDFT predictions in the $\delta$ and $\beta$ phases for a perfect crystalline system and a lattice with Ti vacancies, respectively. Background colors identify the range of stability of the various structural phases.
}\label{fig:Tc}
\end{figure}

At ambient conditions, Ti crystallizes in a hexagonal close-packed structure (Ti-$\alpha$ phase) stable until 8 GPa. On increasing the pressure, a sequence of different phases is found: Ti-$\omega$ (8-100 GPa), Ti-$\gamma$ (100-140 GPa), Ti-$\delta$ (140-243 GPa), which is a distorted bcc crystal, and lastly the Ti-$\beta$ phase (bcc), stable from 243 GPa onwards (See Fig.~\ref{fig:Tc}).
Titanium is superconducting starting from the $\omega$-phase with the highest temperature (\Tc\ $>$ 20K) occurring at the boundary between the $\delta$ and $\beta$ phases. 
The experimental critical temperatures agree reasonably well with the computational predictions under the assumption of a conventional electron-phonon mechanism for pressures below 120 GPa, {\em i.e.} for the $\omega$ and $\gamma$ phases; therefore, in this work we have focused on the $\delta$ and $\beta$ phases of Ti.%, where predictions markedly disagree with experiments. 

We have performed structural relaxations and electron-phonon coupling calculations between 120 and 300 GPa, (at 120, 150, 200, 250 and 300 GPa),
for the $\delta$ and $\beta$ phases. The calculated \Tc's are shown in Fig.~\ref{fig:Tc} where they are compared with the experimental data from Refs.~\onlinecite{Zhang2022_RecordHiTcElementSCinTi_NatComm2022} and~\onlinecite{Liu_deltaPhase_Ti_200GPA_PhysRevB.105.224511}. As one can see, while in the range between 120 and 150 GPa, the calculated \Tc\ is in good agreement with experiments\cite{note1}, beyond 150 GPa it rapidly drops departing from the experimental values, which stay almost constant: at 300~GPa  the predicted critical temperature \Tc=6.9~K is three times smaller than the corresponding experimental estimate of 20~K. Our results are consistent with the calculations in Ref.~\onlinecite{Zhang2022_RecordHiTcElementSCinTi_NatComm2022}, but disagree with those of Ref.~\onlinecite{Liu_deltaPhase_Ti_200GPA_PhysRevB.105.224511}. In this regard we observe that the "anomalous $\mu^*$ scenario" put forward by Liu {\em et al} is ruled out when estimating $\mu^*$ from first principles. 
A non empirical $\mu^*$ can be defined as that number which, used in the McMillan equation~\cite{AllenMitrovic1983}, yields the \emph{ab initio} value of \tc. From the SCDFT value of \tc\ we can estimate $\mu^*$ to be in the range 0.08-0.14, depending on the phase and pressure. These values are a factor of two lower than that proposed by Liu {\em et al}\cite{Liu_deltaPhase_Ti_200GPA_PhysRevB.105.224511} and totally normal for conventional superconductors. 

%, i.e. a factor of two lower than the $\mu^*$ proposed by Liu {\em et al}\cite{Liu_deltaPhase_Ti_200GPA_PhysRevB.105.224511} and   perfectly "conventional" for conventional superconductors. 

 \begin{figure*}[t] 
\begin{center}
\includegraphics[width=0.7\textwidth]{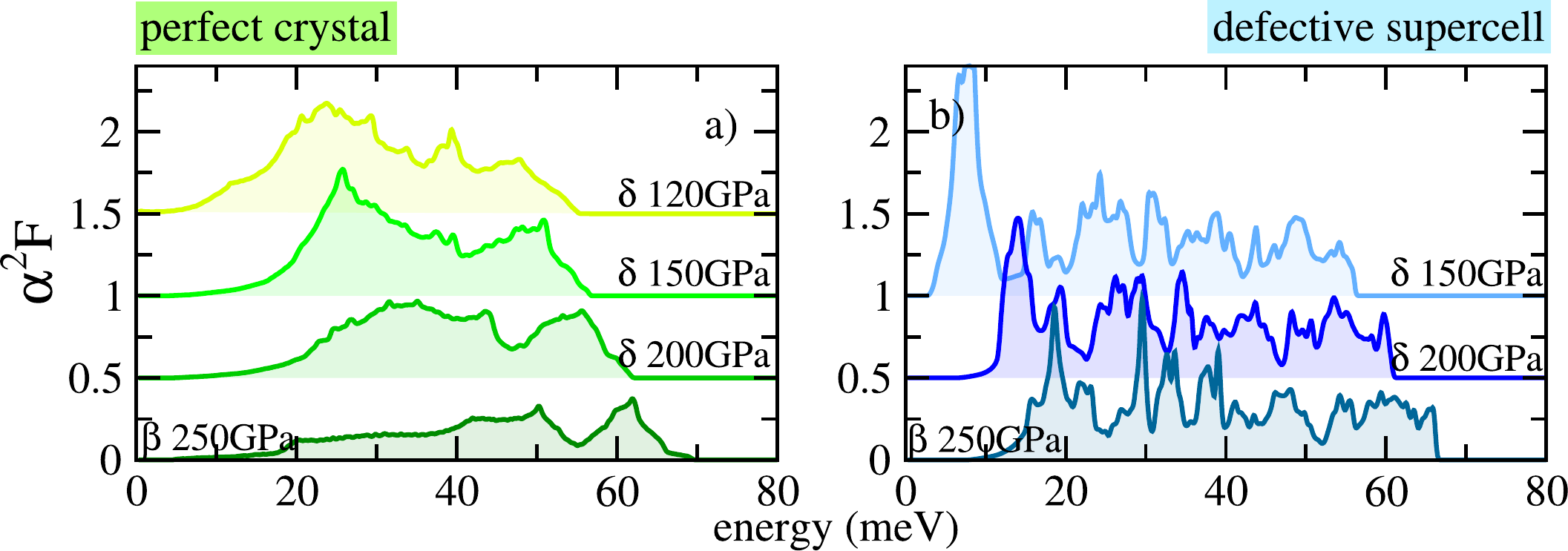} 
\caption{Pressure dependence of the electron-phonon coupling \af\ function in the range 120-250 GPa for the pristine bcc phase (left panel) and the defective bcc phase (right panel) of compressed Ti.
}\label{fig:a2FP}
\end{center}
\end{figure*}

The disagreement we have found between calculations and experiments may indicate an unconventional contribution to the superconducting coupling, e.g., spin fluctuations, as it was suggested in Ref.~\onlinecite{Zhang2022_RecordHiTcElementSCinTi_NatComm2022}. However, we believe this possibility to be unlikely, because (i) the computed Fermi surface consists of poorly nested pockets~\cite{Stewart_UnconventionalSCreview_AdvPhys2017,Annett_UnconventionalSC_1995}, and (ii) the calculated magnetic susceptibility (see Supplemental Information~\cite{Supplemental_Material}) neither exhibits peaks~\cite{Essenberger_SpinFluctuationsTheory_PRB2014,Pellegrini_KO_PRB2023} nor has a strong momentum dependence, which would be necessary to sustain a phase dependent order parameter~\cite{Mazin_Spm_LaOFeAs_PRL2008}.

Another conceivable explanation may be that the crystal phase where high Tc occurs is not the thermodynamically stable bcc one, but some other (possibly metastable) phase formed due to the complex thermodynamical conditions for high pressure synthesis. We have explored this idea by means of  evolutionary crystal structure prediction based on DFT, as implemented in the USPEX code\cite{GLASS2006713, LYAKHOV20131172}. The initial search has been performed at 300~GPa and the best structures have been then relaxed at several intermediate pressures down to 200 GPa. According to our results, at 200~GPa the ground-state structure is indeed the bcc ($Im3m$ space group) phase, while the second-lowest local minimum exhibits a $C2/c$ space group, and an enthalpy 90 meV/atom higher than the $Im3m$ phase\cite{foot_crystal}.
The $C2/c$ phase can be viewed as a distorted bcc phase, and indeed the calculated \Tc\ is virtually identical to the bcc one. 
We thus conclude that none of the phases close to stability appear to be significantly different from the bcc, nor they appear to have significantly stronger superconducting pairing to justify a larger \Tc.

A hint for solving this puzzle has come to us by studying the pressure dependence of \Tc\ within McMillan-Hopfield theory\cite{Supplemental_Material,Th:Hopfield_PR_1969,McMillanTC}. This approach combines the McMillan formula for \Tc\ with the Hopfield expression for the electron-phonon coupling $\lambda=N_F I^2/M\omega^2$, where $M$ is the atomic mass, $\omega$ is a characteristic phonon frequency and $I$ is the electron-phonon deformation potential~\cite{KhanAllen_DeformationPotential_PRB1984,MarchJones_Book2_1973}. 
% therefore assuming that a given superconductor with atomic mass $M$ may be modelled by a single characteristic phonon frequency $\omega$ and an electron-phonon deformation potential~\cite{KhanAllen_DeformationPotential_PRB1984,MarchJones_Book2_1973}, $I$.  
For fixed values of $I$ and the density of states (DOS) at the Fermi level (N$_F$), one observes that there is an optimal value of  $\omega$ which maximises \Tc. For bcc-Ti at 300 GPa this optimal value turns out to be $\simeq$12~meV which would raise \Tc\ to 16K (it should be noted that the actual computed \omlog\ is 34 meV, three times larger than the optimal frequency). This analysis (see~\onlinecite{Supplemental_Material} for more details) suggests that the observed high \tc\ in titanium under pressure might result from an extrinsic effect able to lower the phonon vibrational frequencies. We have explored this possibility by considering the role of structural defects, and more specifically Ti vacancies, which may likely appear in real samples owing to the complex dynamics of high pressure synthesis.
 
\subsection{Defective $\delta$ and $\beta$ phases of $\text{Ti}$ at high pressure}\label{sec:phases}

We have modeled a single vacancy in the high pressure phases of Ti ($\beta$ or bcc-like $\delta$) by removing a Ti atom from a $2\times2\times2$ supercell of the conventional bcc unit cell. For this large simulation cell (containing 15 atoms) we have carried out structural relaxation and computed the electronic, vibrational and superconducting properties as a function of pressure. The results for Tc\ are shown in Fig.~\ref{fig:Tc}. The electron-phonon couplings as a function of pressure are presented in Fig.~\ref{fig:a2FP}. 

\begin{figure*}[t] 
\includegraphics[width=0.7\textwidth]{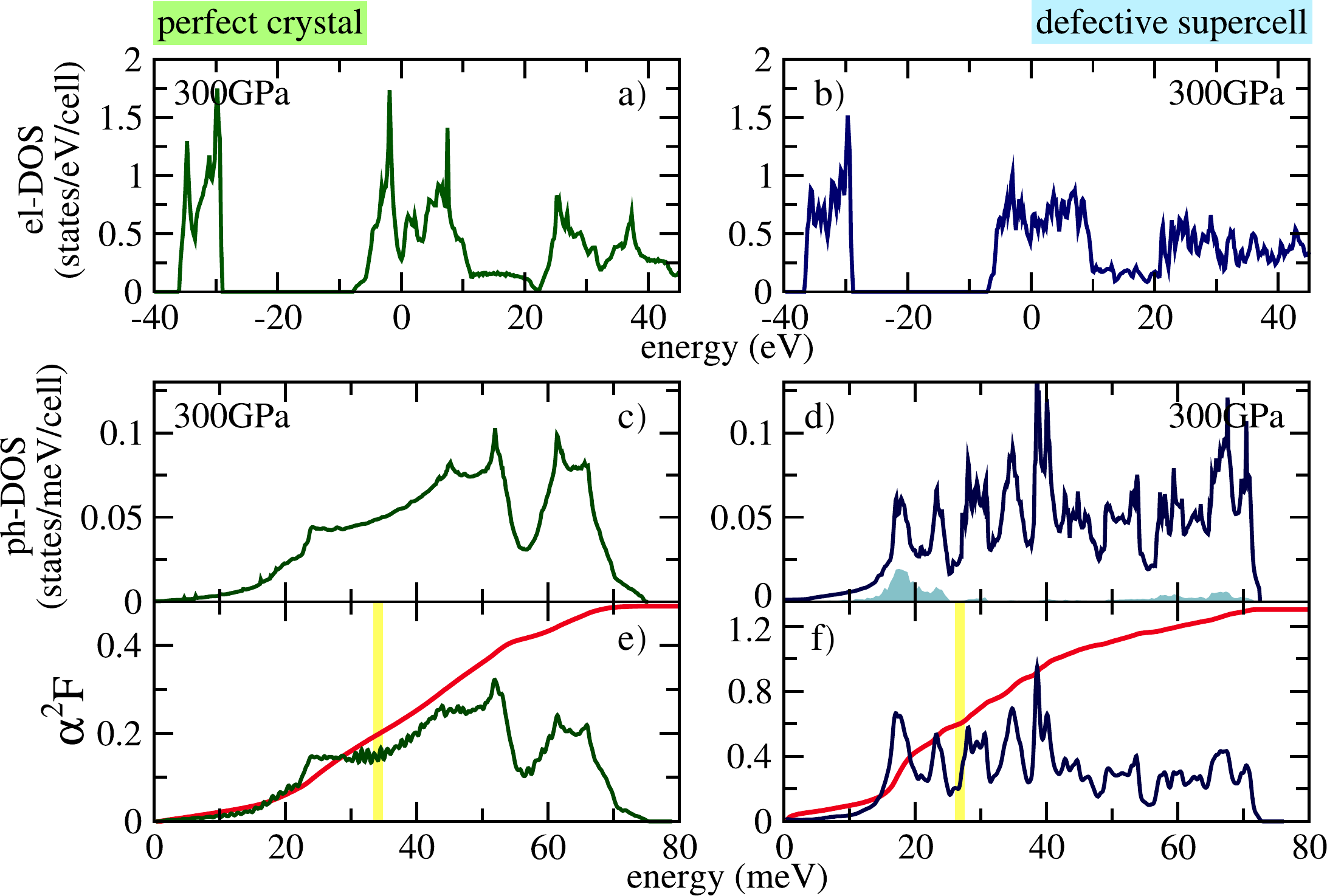} 
\caption{Electronic and vibrational properties of compressed Ti at 300 GPa in the pristine bcc phase (left panels) and defective bcc phase (right panels). (a) and (b): electronic density of states (DOS). (c) and (d): phonon DOS. The blue shaded area in (d) highlights the contribution from a single Ti atom with low bonding coordination.
(e) and (f): electron-phonon \af\ function and its integration curve up to $\lambda$. The yellow vertical lines indicate the value of the logarithmic averaged phonon frequency \omlog. All extensive quantities are normalised to a single Ti unit cell.}\label{fig:data}
\end{figure*}

For the sake of simplicity, we will discuss primarily the results at 300~GPa. The relaxed supercell (crystal structure given in the Supplemental Information~\cite{Supplemental_Material}) is dynamically stable on a  $4\times4\times4$ phononic ${\bf q}$-grid. The corresponding electronic DOS, phononic DOS and electron-phonon coupling at 300~GPa are shown in Fig.~\ref{fig:data}, where they are compared with the same quantities calculated for the ideal (non defective) phase at this pressure. 

From the comparison of the behavior between panels (a) and (b) of Fig.~\ref{fig:data} it is clear that the significant distortion of the lattice causes a smoothing of several sharp features in the electronic DOS. Most importantly for superconductivity, it fills up a sharp dip in the vicinity of the Fermi level increasing the DOS at the Fermi energy from about 0.3 states/eV/atom to about 0.48 states/eV/atom.

However, the most striking effect of the vacancies is  observed in the lattice dynamics, where we find a sensible reduction of the average phonon frequency with respect to the fully symmetric structure. This can be appreciated by comparing the phonon DOS in Figs.~\ref{fig:data} (c) and (d). A detailed analysis of the dynamical matrices and their atomic decomposition in the defective bcc phase shows the existence of localized (i.e. involving few-atom) modes, which strongly contribute to the lowest DOS peak [marked as a blue area in Fig.~\ref{fig:data}(d)]. 
%This effect is expected in disordered phase and should provide additional coupling for the superconducting pairing. 
Indeed, the combined effect of the increased N$_F$ and softer phonon modes leads to a much larger $\alpha^2F$ coupling function [Fig.~\ref{fig:data} (f) vs (e)], which integrates to a large coupling constant $\lambda=1.3$ (to be compared with $\lambda=0.49$ for the ideal bcc phase). 
The predicted \Tc\ value for the defective phase computed within SCDFT is 26~K, which is extremely close to the experimental measurement at this pressure. Note that %in spite of the fact that  \omlog\ is larger than the optimal frequency of the unperturbed system,
the McMillan-Hopfield analysis now applied to the defective system gives an optimal phonon frequency of 24~meV, close to the calculated value of \omlog=28~meV, which suggests we may have achieved the highest \Tc\ for Ti in this lattice type at 300 GPa. 

By applying the same computational approach at lower pressures to the defective $\beta$ and $\delta$ phases, we have obtained the critical temperatures reported as blue symbols in Fig.~\ref{fig:Tc}. 
We find that our defective model results dynamically unstable below 120~GPa, which however bears no relevance since Ti undergoes a phase transition to the $\gamma$ phase at 148 GPa.
Interestingly, in the higher pressure range, from 150 to 300 GPa, we predict a \Tc\ within 25-26 K, rather constant as a function of pressure, in excellent agreement with the experimental observations.

Tab.~\ref{tab:elph_data} summarizes the comparison between the calculated values of the electron-phonon coupling ($\lambda$), \omlog, and \Tc\  for the defective and ideal bcc phases of titanium. 

\begin{table}
\begin{tabular}{| l | l | l | c | c | c |}
\hline
 P (GPa) &\multicolumn{2}{|l|}{Structure}  & $\lambda$ & \omlog\ (meV) & \Tc\ (K) \\
\hline\hline
120 & $\delta$ & dist. bcc & 1.32 & 21.5 & 22.6 \\ 
150 & $\delta$ & dist. bcc & 0.98 & 28.7 & 20.6 \\
200 & $\delta$ & dist. bcc & 0.77 & 32.7 & 16.9 \\
250 & $\beta$  & bcc       & 0.49 & 34.0 &  9.4 \\
300 & $\beta$  & bcc       & 0.49 & 34.1 &  6.9 \\
\hline\hline
150 & def-$\delta$ &def. dist. bcc & 2.73 & 11.6 & 26.7 \\
200 & def-$\delta$ &def. dist. bcc & 1.39 & 23.1 & 24.4 \\ 
250 & def-$\beta$  &def. bcc   & 1.14 & 28.7 & 25.2 \\
300 & def-$\beta$  &def. bcc   & 1.24 & 28.4 & 26.3 \\  \hline
\end{tabular}
\caption{Electron-phonon coupling parameters ($\lambda$, \omlog) and SCDFT critical temperatures (\tc) as a function of pressure (P) for the bcc, distorted (dist. bcc),  defective (def. bcc) and defective distorted (def. dist. bcc) structural phases of Ti studied in this work.}\label{tab:elph_data}
\end{table}

 %placing it about 40 meV/atom higher than the single vacancy defect in the Im3m phase

% \begin{table}
% \begin{tabular}{l|| c | c | c | c | c}
%     & $N_F$  & \omlog\ & $\lambda$ & $\mu$ & \Tc\ \\
%     &(st./eV/cell/spin) &  (meV) & &  &  (K) \\
% \hline 
% bcc ($\beta$-phase) &  0.18 & 34.0 & 0.49 & 0.16 &  6.1  \\
% defective bcc       &  0.26 & 24.2 & 1.30 & 0.24 & 28.0 
% \end{tabular}
% \caption{Summary of electronic and vibrational parameters computed for the two lattices under consideration: the symmetric  and the defective bcc-Ti. All parameters are computed from first principles at the theoretically relaxed lattice. Tc is computed with density functional theory for superconductors. }\label{tab:data}
% \end{table}

\section{Discussion and Conclusions}
We have considered the peculiar case of high-pressure superconductivity at 20~K in bcc Ti recently reported by~Zhang and collaborators\cite{Zhang2022_RecordHiTcElementSCinTi_NatComm2022} and Liu and coworkers \cite{Liu_deltaPhase_Ti_200GPA_PhysRevB.105.224511}. At 300~GPa this measurement is apparently in contrast with the predictions from \emph{ab initio} theory of electron-phonon superconductivity, which yields a significantly lower critical temperature of approximately 6~K. It was suggested that this anomaly could be an indication of an unconventional paring mechanism\cite{Liu_deltaPhase_Ti_200GPA_PhysRevB.105.224511}. 

In our work we present an alternative explanation for this striking discrepancy between theory and experiments. 
Driven by the observation that phonon frequencies in titanium are extremely high, far from optimal values for superconductivity compared to the deformation potential of Ti under pressure, we have formulated the hypothesis that structural defects might be able to optimize the phonon  frequencies with respect to the electron-phonon coupling, possibly leading to a higher \tc. 
To verify this hypothesis we have investigated the effect of vacancies on the structural, electronic, dynamical and superconducting properties of Ti by means of accurate first-principles theory (without any semi-empirical parameters).
We predict that Ti vacancies can substantially increase the superconducting critical temperature under pressure, up to values as high as 26~K, which is close to the experimental estimate. 
%Although our proposed mechanism  quantitatively account for the extremely high observed \Tc's in Ti under pressure, 
Some additional considerations are in order:

(i) While the doping level we have considered is relatively high (of the order of 7\%), we expect the described mechanism to be effective also at significantly lower defect density. In fact, given the relatively long coherence length of the system (about 6~nm as estimated from the superconducting gap function and the Fermi velocity~\cite{Carbotte_RMP1990}), even if a relatively small fraction of the sample volume (around the impurities) supported high \tc, these regions could easily  couple within the coherence length of the condensate.

(ii) The effect of impurities becomes increasingly less relevant at lower pressures. In fact, while at 300~GPa pure bcc Ti has stiff phonons (\omlog=35~meV), which can be effectively softened by vacancies (\omlog=28meV), at low pressures phonons are already soft and not significantly affected. %Therefore defects are ineffective.

(iii) 
%In metals under pressure other structural defects can be formed.
We think that the proposed \tc\ enhancement mechanism should be quite general and effective also for other types of defects (dislocations, grain boundaries, interstitials, etc.) and in other superconducting materials.
The possibility of this effect to occur can be assessed through an analysis based on McMillan-Hopfield theory, as shown in this work. However, one should consider that the underlying assumption of this method, i.e., that the relevant material parameters $N_F$ and $I$ are little affected by structural defects, appears reasonable for simple high-symmetry metals, but is quite unlikely for complex covalent compounds.

In conclusion we have proposed an explanation for the high critical temperature in bcc Ti at high pressure, a puzzling phenomenon recently observed. Our explanation is based on the assumption that the system should locally develop Ti vacancies in the bcc lattice. We predict that these defects effectively lower the vibrational frequency and increase the DOS at the Fermi energy, leading to a sizable increase of \Tc. We find excellent agreement between measured and computed \Tc\ values in the 150-300 GPa range. 
Our prediction could find an experimental validation by observing that high quality samples have significantly lower \tc. On the other hand, increasing the number of defects should increase \tc\ only very slightly (by a few Kelvins).

%Our prediction has several experimental implications which can be used for its validation:
%Our prediction could be motivate different experimental verifications: 
%on one side high quality crystals or carefully conducted pressure experiments should display lower \Tc's; on the other side  \tc\ could be further (slightly) increased by increasing the phonon softening by introducing more defects.

%%%%%%%%%%%%%%%%%%%%%%%%%%%%%%%%%%%%%%%%%%%%%%%%%%%%%%%%%%%%%%%%%%%%%%%%%%%%%
\section*{Acknowledgements}

G. P. acknowledges financial support from the Italian Ministry for Research and Education through PRIN-2017 project ``Tuning and understanding Quantum phases in 2D materials - Quantum 2D" (IT-MIUR Grant No. 2017Z8TS5B) and funding from the European Union - NextGenerationEU under the Italian Ministry of University and Research (MUR) National Innovation Ecosystem grant ECS00000041 - VITALITY - CUP E13C22001060006.
L.B. acknowledges support from Fondo Ateneo Sapienza 2019-21, and funding from the European Union - NextGenerationEU under the Italian Ministry of University and Research (MUR), “Network 4 Energy Sustainable Transition - NEST” project (MIUR project code PE000021, Concession Degree No. 1561 of October 11, 2022) - CUP C93C22005230007. 
%\bibliography{paper}

\begin{thebibliography}{99}
\bibitem [{not()}]{note1}%
  \BibitemOpen
  \href@noop {} {}\bibinfo {note} {Our method doesn't involve any adjustable
  parameter like the $\mu*$ used in most Eliashberg theory applications,
  therefore the agreement is not enforced by any fitting
  procedure.}\BibitemShut {Stop}%
\end{thebibliography}
%

\end{document}